\documentclass[AMA,LATO1COL]{WileyNJD-v2}

\articletype{Methodological}%

\received{26 April 2016}
\revised{6 June 2016}
\accepted{6 June 2016}

\usepackage{graphicx}
\usepackage{multirow}
\usepackage{tabularx}
\usepackage{amsmath}
\usepackage{listings}
\usepackage{color}

\definecolor{dkgreen}{rgb}{0,0.6,0}
\definecolor{gray}{rgb}{0.5,0.5,0.5}
\definecolor{mauve}{rgb}{0.58,0,0.82}

\newtheorem{tiarule}{Rule}

\begin{document}

\title{An approach for Test Impact Analysis on the Integration Level in Java programs}

\author{Muzammil Shahbaz}

\address{\orgdiv{Thales Under Water Systems}, \orgaddress{\country{United Kingdom}}}

\corres{\email{muzammil.shahbaz@uk.thalesgroup.com}}

\newcommand{\jvmsniffer}{\textit{JVMSniffer} }
\newcommand{\jtia}{{jTIA}}

\abstract[Summary]{Test Impact Analysis is an approach to obtain a subset of tests that are impacted by the code changes. This approach is mostly applied to unit testing where the link between the code and its associated tests is easy to obtain. On the integration level, however, it is not straightfoward to find such a link programatically, especially when the integration tests are held into separate repositories. We propose an approach for selecting integration tests based on the runtime analysis of code changes to reduce the test execution overhead. We provide a set of tools and framework that can be plugged into existing CI/CD pipelines. We have evaluated the approach on a range of opensource Java programs and found $\approx$50\% reduction in tests on average, and above 80\% in a few cases. We have also applied the approach to a large-scaled system currently in production and found similar results.}


\maketitle

\section{Introduction}\label{sec:introduction}

	Continuous Integration/Delivery (CI/CD)~\cite{Humble2010} is a modern way of building and releasing software in a shorter development lifecycle. Developers continuously implement small changes into their code and merge back to the main branch of the repository as often as possible and many times a day. These changes are validated by the automated system of building, testing, and deployment of software, which makes the integration easier and delivery quicker. Running automated integration tests is a vital phase of CI/CD that makes sure that the new changes to the code does not break the software or acceptance criteria whenever new merges  are committed to the main branch. Moreover, they also serve as a regression test suite, verifying that no bugs are introduced into the existing behavior by the new changes. 

\begin{figure*}
 \centering
  \includegraphics[scale=0.4]{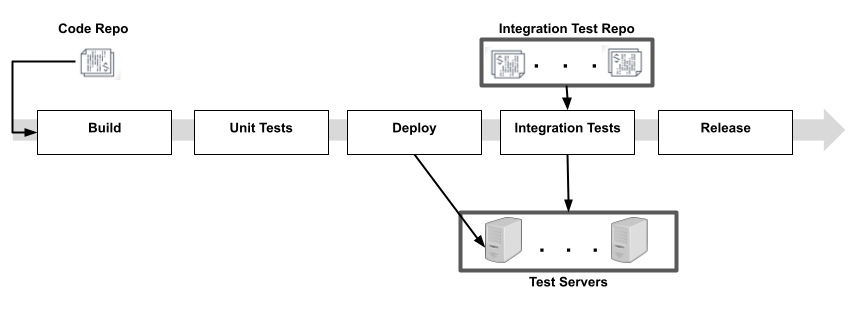}
\caption{Typical stages of a CI/CD pipeline.}
\label{fig:pipeline}
\end{figure*}

Figure~\ref{fig:pipeline} shows a typical CI/CD pipeline that how a software component is moved across various stages in the development lifecycle before it is released. The pipeline is triggered when some code changes are committed into a repository of a version control system. The code goes to the build phase where it is compiled, followed by running the unit tests. Given there are no failures in the build, the component moves to the next phase where it is deployed in stagging or a test server. Thereby, the integration or acceptance tests are checked out from the test repository and run on the component along with the other components in the system collectively. When all the integration tests are passed, the component is tagged or promoted to release.  

\subsection{Motivation}

As software matures with time, the integration test-suite tends to grow to the extent that running all tests takes a considerable amount of time. 
This is especially true when the software is delivered via iterative process, where the first release might contain a minimum functionality, but its codebase grows manifolds due to adding new features and fixing defects. This increases the rate of code changes, for instance, Google's code churn rate --- a commit every second on  average --- produces 800K builds with 150 Million tests runs daily~\cite{Memon2017}. Even with the company's massive compute resources, it is not cost effective to test each code commit individually at this rate. In practice, developers delay their merges until tests are completed in the pipeline for each of the prior changes. This eventually slows down the whole pipeline process and diminishes the benefit of automation. Developers tackle this situation by either disabling some tests, or deferring them to the very end of the release cycle, which again results in lower productivity and quality.


\subsection{Test Impact Analysis}

The basic principle of \textit{Test Impact Analysis (TIA)}~\cite{Hammant2017}\cite{vstudio_magazine}\cite{Peng2020} is to determine the subset of tests that is impacted by the code change. Thus, running only the subset results into faster execution of the pipeline.
 
Finding the subset of impacted tests is easy on a component level where code and unit tests are kept in the same repository. 
Modern IDEs and tools for structural coverage~\cite{Molnar2017} can efficiently compute call graphs of unit tests either statically or dynamically. These call graphs can trace the tests back to the methods changed and then only those unit tests can be selected to run. 

On the integration level, however, is a different story. The integrated tests cover the whole system and  execute many components at once or in a particular order. There might not be a direct relation between an integration test and a method changed in one of those components. This is a typical case of \textbf{microservices}~\cite{Indrasiri2018}, where a monolithic application is replaced by a suite of small services that are built independently. They are implemented in different repositories, run on different servers and communicate via different network protocols. 
Thus there is no easy way of determining the (indirect) dependencies between the tests and the new commits in various components.

We propose an approach for test selection that combines static and dynamic analysis of Java components irrespective of their organization into repositories or system topology. We have implemented the approach into existing CI/CD pipelines, so that it only runs a subset of tests required to validate the code being committed without losing quality, i.e., the outcome of the tests that are not selected will not be affected by the changes. Trivially, it results into faster execution of the pipeline because for a given code commit, our approach selects and runs only the relevant tests required to validate that commit. 

The rest of the paper is organized as follows. Section~\ref{sec:approach} explains the approach with formal settings. Section~\ref{sec:exp} presents the experimental evaluation. Section~\ref{sec:related_works} discusses the existing works related to this topic. Section~\ref{sec:conclusion} concludes the paper with a note on future works.

\section{Methodology}
\label{sec:approach}

\subsection{Overview}
\label{sec:overview}
In order to perform the impact analysis, we need to understand the relationship between the existing tests and the source code. Our approach is to ``sniff'' the method invocations at the test execution runtime. At the end of testing, we produce a map that links each test with the associated list of methods invoked. 
For instance, we have a map of three tests and the methods they call when executed:

\begin{lstlisting}
	map(T1) = {A.m1, B.m2, C.m1}
	map(T2) = {A.m2, B.m1}
	map(T3) = {B.m2, C.m2}
\end{lstlisting}

That means the test \texttt{T1} calls the method \texttt{m1} of class \texttt{A}, \texttt{m2} of class \texttt{B} and \texttt{m1} of class \texttt{C}. Similarly, the test \texttt{T2} calls the method \texttt{m2} of class \texttt{A} and \texttt{m1} of class \texttt{B}, and so on. 

We store this map in the integration tests repository. In the next code commit, we compute the difference with the previous commit to analyze code changes. Thereby, we determine which tests are actually impacted by looking at this map and the methods that are changed in the last commit. Suppose the next commit changes the method \texttt{m2} of the class \texttt{B}, then we perform the lookup into the map and pick tests \texttt{T1} and \texttt{T3}; the only tests that are impacted. We leave \texttt{T2} as it is not relevant for this change. 

Our approach is \textit{safe}~\cite{Yoo2012} in the sense that the set of selected tests contains all tests whose behavior may have been affected by the change in the methods. 

\begin{figure*}[t]
	\centering
	\includegraphics[scale=0.4]{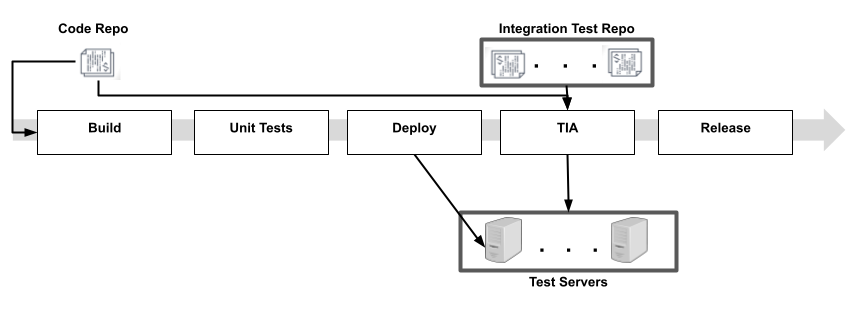}
	\caption{Stages of the CI/CD Pipeline with TIA adjusted.}
	\label{fig:pipeline_tia}
\end{figure*}

Figure~\ref{fig:pipeline_tia} shows the pipeline with TIA adjusted. 
We define the concepts used in the approach in the following section, and then explain the implementation stages of the approach.

\subsection{Formal Settings}
\label{sec:formal}

We assume the source code consists of classes and each class has a set of implemented methods. We consider explicit constructors as methods. We also assume a set of tests that call a subset of those methods when they are executed.

Let $M$ be the set of all methods and $T$ be the set of all tests. We assume that $M$ and $T$ are non-empty sets, i.e., there are one or more methods in the source code and one or more tests that execute a subset of those methods. For this purpose, we define a function $\lambda : T \times M \longrightarrow \{1, 0\}$, such that $\lambda(t, m) = 1$ iff the method $m$ is executed by the test $t$, and $\lambda(t, m) = 0$, otherwise. We also use a \textit{map} for each test and the methods it executes defined as $T \longrightarrow 2^{M}$ such that each $t \in T$ maps to a set $\{m_1, \cdots, m_n\} \subseteq M$ where $\lambda(t, m_i) = 1$ for $1 \leq i \leq n$. 



A method declaration has six components: name, arguments, return type, modifiers, exception list and an implementation body~\cite{Java8_specs}.

A method is called \textbf{\textit{modified}} if the method is 1) changed, 2) removed, or 3) added. Each of these types are explained below. 

\subsubsection{Changed}
\label{sec:changed}

A method is changed if its modifiers, exception list or implementation body is changed. 
Thus, we create the following rule to track method changes.

\begin{tiarule}
	\label{rule:changed}
	A method is marked \textit{modified} if any of the following changes occur
		
		\begin{itemize}
			\item  implementation body, e.g., adding/removing/changing statements.
			\item modifiers, e.g., adding/removing access modifiers (including \texttt{static}) or keywords like  \texttt{synchronized}, \texttt{throws} or \texttt{abstract}.
			\item exception list, e.g, adding/removing checked exceptions.
		\end{itemize}
	\end{tiarule}

Changes to return types do not pose a particular challenge to our methodology. This is because the return statement in the implementation body is changed almost every time the return type is changed. The only exception to changing the return type without changing the return statement is the \textit{covariant} return type~\cite{Java8_specs}, which does not impact the semantic change without changing the implementation. 

Changes to a method name or arguments force changes into other methods which call the changed method. In this case, the other methods will be marked as modified as per Rule~\ref{rule:changed}. 

If a method has not been called by any other method, but changing its name or arguments cause it to override/overload an existing method, then it can alter the behavior of the program. Therefore, any changes to the name or arguments are considered as the method is removed and a new method is added. Both of these cases are covered in the next sections.

\subsubsection{Removed}
\label{sec:removed}

A method is removed if its whole declaration is deleted, or changes to its name or arguments occur. 
Thus, we create the following rule to track method removals.

\begin{tiarule}
\label{rule:removed}
A method is marked \textit{modified} if any of the following changes occur

\begin{itemize}
	\item declaration is removed.
	\item name is changed.
	\item arguments (including number, order or types) are changed.
\end{itemize}

\end{tiarule}

All methods having explicit references to a removed method would require changes to its implementation. Those methods will also be marked \textit{modified} as per Rule \ref{rule:changed}.
Removing a method may also cause some tests to be invalid if they have an explicit reference to the method. Any adjustments to tests due to a source code change is not in the scope of our methodology.

\subsubsection{Added}

Any new methods likely result into changes to existing methods or tests. However, if the new method  overrides or overloads an existing method, it may alter the behavior of tests without changing the existing code.

If the new method overrides an existing method, the new method may be called in lieu of the existing method due to dynamic binding\footnote{Version 1 in Figure \ref{fig:prog_versions} provides an example of such a case.}. In this case, we mark the existing method as \textit{modified}. We create the following rule to cover this case.

\begin{tiarule}
	\label{rule:override}
	If a method $\alpha$ is a new method such that it overrides an existing method $\beta$, then $\beta$ is marked \textit{modified}.
\end{tiarule}

If the new method overloads an existing method such that it matches the name, number, order and type of arguments with another method in the class hierarchy, then it can change the static binding of the method in the other class\footnote{Version 2 in Figure~\ref{fig:prog_versions} provides an example of such a case.}. This is because the binding of such a method depends upon the compile-time types of the arguments~\cite{Java8_specs}. Therefore, all methods in the class hierarchy that has the same name, number, order, and type of arguments are marked as \textit{modified}. We create the following rule to cover this case.

\begin{tiarule}
	\label{rule:overload}
	If a method $\alpha$ is a new method such that it matches the name, number, order and (super or sub) type of arguments with a method $\beta$ in any class in the hierarchy, then $\beta$ is marked \textit{modified}.
\end{tiarule}

\begin{figure}
	\begin{minipage}[t]{.3\linewidth}
		\begin{tabular}[t]{l}
			\hline
			\textbf{Program} \\
			\hline
			\begin{lstlisting}
class A {
	void foo() {
	}
	void bar(Object obj) {
	}
}

class B extends A {
	void foo() {
	}
}
			\end{lstlisting}
		\end{tabular}
	\end{minipage}
	\hfill
	\begin{minipage}[t]{.7\linewidth}
		\raggedleft
		\begin{tabular}[t]{l | l | l}
			\hline
			\multicolumn{3}{c}{\textbf{Tests}} \\
			\hline
			\begin{lstlisting}
// T1
test1() {
	A a = new A();
	a.foo();		
}
			\end{lstlisting}
			
			&
			
			\begin{lstlisting}
// T2
test2() {
	A a = new B();
	a.foo();		
}
			\end{lstlisting}			
			
			&
			
			\begin{lstlisting}
// T3
test3() {
	A a = new B();
	a.bar("hello");		
}	
			\end{lstlisting} \\
			
			\hline
			
			\multicolumn{3}{c}{\textbf{Mappings}} \\
			\hline
			
			\begin{lstlisting}
// map(T1)
{ A.A(), A.foo() }
			\end{lstlisting}
			
			& 
			
			\begin{lstlisting}
// map(T2)
{ B.B(), A.A(), B.foo() }
			\end{lstlisting}
			
			&
			
			\begin{lstlisting}
// map(T3)
{ B.B(), A.A(), 
	A.bar(Object) }	
			\end{lstlisting}

		\end{tabular}
	\end{minipage}
	\caption{Example: Original program (left) and its associated tests with mappings (right).}
	\label{fig:original_prog_tests}
\end{figure}

\begin{figure}
	\begin{tabular}[t]{l|l|l|l}
		\hline
		\textbf{Version 1: Overriden} & \textbf{Version 2: Overloaded} & \textbf{Version 3: Changed} & \textbf{Version 4: Removed} \\
		\textit{\texttt{B.bar(Object)} added} & \textit{\texttt{A.bar(String)} added} & \textit{\texttt{B.foo()} changed} & \textit{\texttt{A.foo()} removed} \\
		\hline
		
		\begin{lstlisting}
class A {
	void foo() {
	}
	void bar(Object obj) {
	}
}


class B extends A {
	void foo() {
	}
	void bar(Object obj) {
	}
}
		\end{lstlisting}
		
		&
		
		\begin{lstlisting}
class A {
	void foo() {
	}
	void bar(Object obj) {
	}
	void bar(String text) {
	}
}
class B extends A {
	void foo() {
	}
	void bar(Object obj) {
	}
}
		\end{lstlisting}
		
		&
		
		\begin{lstlisting}
class A {
	void foo() {
	}
	void bar(Object obj) {
	}
	void bar(String text) {
	}
}
class B extends A {
	void foo() {
		bar("hello");
	}
	void bar(Object obj) {
	}
}
		\end{lstlisting}
		
		&
		
		\begin{lstlisting}
class A {
	void bar(Object obj) {
	}
	void bar(String text) {
	}
}


class B extends A {
	void foo() {
		bar("hello");
	}
	void bar(Object obj) {
	}
}
		\end{lstlisting} \\
		\hline
		
		\multicolumn{4}{c}{\textbf{Modified Methods}} \\
		\hline
		\texttt{A.bar(Object)} & \texttt{A.bar(Object)} & \texttt{B.foo()} & \texttt{A.foo()} \\
		& \texttt{B.bar(Object)} &            &            \\
		\hline
		
		\multicolumn{4}{c}{\textbf{Selected Tests}} \\
		\hline
		\texttt{T3} & \texttt{T3} & \texttt{T2} & \texttt{T1} \\
		\hline
		
		\multicolumn{4}{c}{\textbf{New Mappings}} \\
		\hline
		\begin{lstlisting}
// map(T1)
{ A.A(), A.foo() }
// map(T2)
{ B.B(), A.A(), B.foo() }

// map(T3)
{ B.B(), A.A(), 
	B.bar(Object) }	
		\end{lstlisting}
		
		& 
		
		\begin{lstlisting}
// map(T1)
{ A.A(), A.foo() }
// map(T2)
{ B.B(), A.A(), B.foo() }

// map(T3)
{ B.B(), A.A(), 
	A.bar(String) }	
		\end{lstlisting}
		
		&
		
		\begin{lstlisting}
// map(T1)
{ A.A(), A.foo() }
// map(T2)
{ B.B(), A.A(), B.foo(), 
	A.bar(String) }
// map(T3)
{ B.B(), A.A(), 
	A.bar(String) }	
		\end{lstlisting}
		
		&
		
		\begin{lstlisting}
// map(T1)
N/A - T1 cannot execute
// map(T2)
{ B.B(), A.A(), B.foo(), 
	A.bar(String) }
// map(T3)
{ B.B(), A.A(), 
	A.bar(String) }	
		\end{lstlisting}

	\end{tabular}%
	\caption{Example: Illustration of four versions of the original program in Figure~\ref{fig:original_prog_tests} and test selection with new mappings in each iteration.}
	\label{fig:prog_versions}
\end{figure}

\subsection{Example}
\label{sec:example]}
We explain the approach with the help of the example in Figure \ref{fig:original_prog_tests}. The original program consists of class $ A $ and its subclass $ B $. Three tests, $ T1 $, $ T2 $ and $ T3 $, and their associated maps of method calls can also be seen. The program goes through four versions of changes as shown in Figure \ref{fig:prog_versions}. The figure also shows ``Modified Methods'' in each version, ``Selected Tests'' based on those methods, and ``New Mappings'' calculated after running those tests. We have shown the map of all tests in each iteration for completeness. 

Version 1 adds a new method \lstinline!B.bar(Object)! that overrides \lstinline!A.bar(Object)!, which marks it a modified method (as per Rule~\ref{rule:override}). We find this method in $map(T3)$ and therefore $ T3 $ is selected. The new map for $ T3 $ is obtained after running the test on Version 1 that now includes \lstinline!B.bar(Object)!. 

Version 2 overloads \lstinline!A.bar(Object)! by adding \lstinline!A.bar(String)!. Note that the new method has the same name, number, order and type of arguments as the original method. The only difference is that the new method uses a subtype of the existing method. This alters the behavior of $ T3 $ which now resolves to the new method (instead of \lstinline!B.bar(Object)!) due to the explicit \texttt{String} type argument (i.e., ``hello'') in the test. We mark \lstinline!A.bar(Object)! and \lstinline!B.bar(Object)! as modified methods (as per Rule~\ref{rule:overload}). The latter is included in $map(T3)$, which was updated after Version 1. Thus, $ T3 $ is selected once again and the new map is updated that now includes \lstinline!A.bar(String)!. 

Version 3 changes the body of \lstinline!B.foo()!, which is therefore a modified method (as per Rule~\ref{rule:changed}). As this is included in $ map(T2) $, $ T2 $ is selected and map is updated consequently. 

Version 4 removes \lstinline!A.foo()!, which is therefore a modified method (as per Rule~\ref{rule:removed}). As this is included in $map(T1)$, $ T1 $ is selected. However, this test cannot be executed because of the explicit reference to the non-existent method, and hence needs revision.

\subsection{Implementation Stages}

\begin{figure}[t]
	\centering
	\includegraphics[scale=0.5]{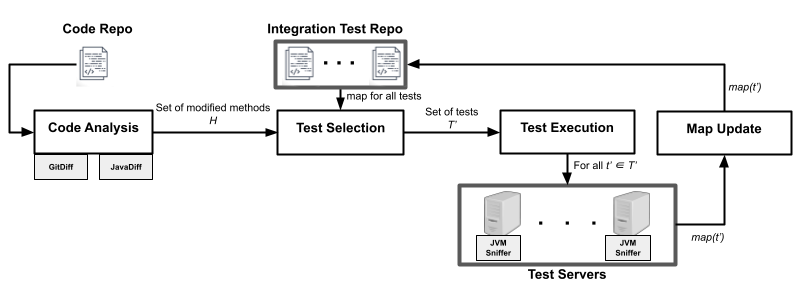}
	\caption{Stages of \jtia: 1) Code Analysis, 2) Test Selection, 3) Test Execution/Map Update}
	\label{fig:tia}
\end{figure}

We have implemented our TIA approach into a pipeline plugin called \jtia, in addition to a set of tools~\cite{TIA2018}\cite{JvmSniffer2018} supporting the different stages of the approach. Figure~\ref{fig:tia} illustrates the stages, which are explained below.

\subsubsection{Stage 1: Code Analysis}

The first stage of \jtia\ performs code change analysis by checking out the component's source code repository. The list of methods $H$ that are modified in the last commit is obtained and passed on to the next stage. 

Our current implementation deals only with GIT repositories for identifying code changes. For this purpose, we have implemented a tool called \textit{GitDiff}~\cite{TIA2018}, which is responsible for managing GIT repositories (e.g., retrieving change logs). 
\textit{GitDiff} retrieves the list of changes in the last commit by executing GIT commands\footnote{\texttt{git diff-tree --no-commit-id --name-only -r HEAD}}
and then parse the output to find the new/changed/deleted Java files. 

Our tool \textit{JavaDiff}~\cite{TIA2018} is responsible for analyzing Java files without regard to formatting, comments, whitespaces, or method ordering. The tool implements the four rules defined in Section~\ref{sec:formal}.
For a changed or deleted Java file, the previous version is checked out\footnote{\texttt{git show HEAD\string^: <file>}} and compared to obtain the list of any methods modified. 
For new Java files, changes in the existing class hierarchy are reported for any overloaded or overridden methods. For this purpose, it uses the call graph analysis of the \textit{Soot} Java optimization framework~\cite{soot}. The current implementation supports up to Java 8.

\subsubsection{Stage II: Test Selection}

The second stage of \jtia\ selects tests based on the list of modified methods from the previous stage. The integration tests repository is checked out and the mapping for each test to its associated methods is obtained. The test that contains a modified method in its mapping is selected. Any test that has no mappings, such as newly added tests, are also selected. This means that all tests are selected when the pipeline is triggered for the very first time. Any test that has been changed since its last execution, or previously failed will have its map deleted, i.e., all changed and failing tests will also be selected. 

Formally, if $H \subseteq M$ is the set of modified methods and $T$ is a set of tests, then the set of selected tests $T' \subseteq T$ is computed in the following way. 
For each test $t \in T$ and a modified method $h \in H$, $t$ is added to $T'$ if 

\begin{enumerate}
	
	\item\label{enum:select2} $map(t) = \emptyset$, or
	\item\label{enum:select1} $h \in map(t)$

\end{enumerate}

\subsubsection{Stage III: Test Execution and Map Update}

The third stage of \jtia\ runs tests in $T'$ one by one on the test servers. For each test $t' \in T'$,   $map(t')$ is recomputed in order to capture any changes after the commit. 

Our tool \jvmsniffer\cite{JvmSniffer2018} is responsible for capturing the method calls for each running test. The tool uses Java Debug Interface API~\cite{JDI} to hook into the target JVM in the debug mode. It provides the ability to trap target events, including method entry and exit events. Therefore, it is able to capture all method invocations on the target JVM objects including types and values of the arguments. It also provides an option to filter method calls based on the package name, which is useful in collecting the information only related to the system under test and avoiding method call traces from third-party or system libraries. 

\jvmsniffer is deployed per JVM on each test server. Before running a test, \jtia\ starts \jvmsniffer on each server that attaches itself to the target JVM. \jtia\ then starts test execution, and \jvmsniffer starts recording all method invocations during the test execution. For each method invocation, it records the name of the package, class, method, its argument list and return type in JVM's representation of type signatures in the format

\begin{verbatim}
 <package>.<class>.<method>(arguments)return_type
\end{verbatim}

When the test is complete, \jtia\ stops all instances of \jvmsniffer and collates the method invocation records from each server. 
Finally, \jtia\ updates the test map with the list of method calls and checks into the integration tests repository. 
Figure~\ref{fig:method-traces} shows the snippet of the list of method calls from one of our case studies.

\begin{figure*}
	
	\begin{verbatim}
		okhttp3.Cache.<init>(Ljava/io/File;JLokhttp3/internal/io/FileSystem;)V
		okhttp3.HttpUrl$Builder$Companion.portColonOffset(Ljava/lang/String;II)I
		okhttp3.internal.cache.DiskLruCache$Companion.create(Ljava/io/File;IIJ)Lokhttp3/internal/cache/DiskLruCache;
		okhttp3.internal.cache.DiskLruCache.<init>(Lokhttp3/internal/io/FileSystem;IIJLjava/util/concurrent/Executor;)V
		okhttp3.internal.cache.DiskLruCache$cleanupRunnable$1.<init>()V
	\end{verbatim}
\caption{Snippet of a method invocation record of the \textit{pragmaNoCache} test from the OkHttp case study in JVM's representation of type signatures.}
\label{fig:method-traces}
\end{figure*}

\section{Evaluation}
\label{sec:exp}

In this section, we evaluate our approach and discuss results, threats to validity of our approach and limitations.

\subsection{Assumptions}

We assume each component in the system is a Java program. Trivially, all components have at least one public class and each of those has at least one method implemented. 
All tests run under identical conditions, i.e, our test environment, resource configuration and external dependencies remain unchanged during the evaluation to produce deterministic results. This means the results are reproducible if we run the same test suite multiple times on the same code base.



\subsection{Case Studies}

The experiments have been conducted on 10 well-known opensource
projects. They are selected from GitHub\footnote{\texttt{https://github.com}} based on their popularity and repository
size to include both popular and huge projects as well as medium and small sized
projects. Table~\ref{tab:case-studies} lists the case studies with their versions and size in terms of lines of code. A short description for each case study is given below. 

\begin{table}
\caption{Case Studies with versions and lines of code in thousands (kLoC).}
\label{tab:case-studies}
\begin{center}
\begin{tabular}{lllll}
\hline\noalign{\smallskip}
\multicolumn{1}{c}{\multirow{2}{*}{\textbf{Case Studies}}} & \textbf{Version} & \multicolumn{3}{c}{\textbf{kLoC}} \\
\multicolumn{1}{c}{}                              &         & total   & src   & test   \\
\noalign{\smallskip}\hline\noalign{\smallskip}
Apache Commons Col.                        & 4.4     & 92      & 51    & 41     \\
Apache Commons Lang                       & 3.6     & 112     & 59    & 53     \\
Eclipse Collections                              & 10.0.0  & 327     & 171   & 156    \\
Netty                                                  & 3.10    & 105     & 87    & 18     \\
RxJava                                               & 2.2.0   & 329     & 134   & 195    \\
Seata                                                 & 0.5.0   & 32      & 24    & 8      \\
OkHttp                                               & 4.0.0   & 47      & 11    & 36     \\
MyBatis                                             & 3.4.0   & 61      & 25    & 36     \\
UML Reverse Mapper                         & 1.4.4   & 2       & 1     & 1      \\
Symja Parser                                     & 1.0.0   & 9       & 7     & 2     \\                
\noalign{\smallskip}\hline
\end{tabular}
\end{center}
\end{table}

\begin{description}
	\item[\textbf{Apache Commons Collections:}] A library from the Apache Software Foundation that extends the Java Collections Framework. 
	\item[\textbf{Apache Commons Lang:}] A library for Java utility classes from the Apache Software Foundation.
	\item [\textbf{Eclipse Collections:}] A collections framework for Java with optimized data structures from the Eclipse Foundation. Only two modules, \textit{forkjoin} and \textit{testutils}, are considered for experiments.
	\item[\textbf{Netty:}] An asynchronous event-driven network application framework for rapid development of maintainable high performance protocol servers and clients.
	\item[\textbf{RxJava:}] A JVM library for composing asynchronous and event-based programs by using observable sequences.
	\item[\textbf{Seata:}] A high-performance distributed transaction solution for microservices architecture.
	\item[\textbf{OkHTTP:}] A performant HTTP/2 client for Android and Java applications.
	\item[\textbf{MyBatis:}] A persistence framework with support for custom SQL, stored procedures and advanced mappings.
	\item[\textbf{UML Reverse Mapper:}] An automatic UML diagrams generator from code.
	\item[\textbf{Symja Parser:}] An implementation of math expressions parsers for Java double values and Apache Commons Math's complex numbers.
\end{description}

\subsection{Research Questions}

\newcommand{\RQone}{
\subparagraph{RQ1: }
Are the selected tests from the proposed approach as effective as the whole test-suite?
}

\RQone

To analyze the effectiveness of our approach, we used mutation testing that provides a reliable metric to measure the quality of a test-suite \cite{Andrews2005}. We used \textit{PIT} mutation testing tool~\cite{Coles2016} in our experiments. The mutants are created based on code coverage, i.e., only the methods covered by the tests are mutated.
PIT generates mutants via number of mutation operators targeting various semantics in Java. The details of each operator can be seen on PIT's website\footnote{http://pitest.org/quickstart/mutators/}. The selected operators in our experiments are

\begin{description}
	\item[\textbf{Default operators:}] Conditional Boundary, Increments, Invert Negatives, Math, Negate Conditionals, Return Values, Void Method Call,\\
	 Empty/False/True/Null/Primitive Returns.
	\item[\textbf{Optional operators:}] Constructor Calls, Inline Constant, Non Void Method Call, Remove Conditionals, Remove Increments.
\end{description}

First, we ran the original program with PIT in order to compute the mutation coverage with the whole test-suite. Then, we changed the tool's configuration to run only the tests selected by our approach and computed the mutation coverage again. We compared the mutation coverage between the two runs by looking at the number of mutants killed. Ideally, the mutation coverage of the selected test-suite should be equal to the mutation coverage of the entire test-suite, otherwise the selected test-suite is not complete with respect to the changes introduced to the program due to mutations.

We used default settings of operators to generate as many mutations possible depending on the size of the project. 
There are multiple reasons for which a mutant can be considered as killed, all of which were considered as fault detection.

\begin{itemize}
	\item The test was executed and failed.
	\item The test was run and caused a memory error, e.g., stack overflow or a heap space error.
	\item The mutation caused an endless loop and terminated due to a predefined timeout (i.e., 3000 msec).
\end{itemize}

%

\newcommand{\RQtwo}{
\subparagraph{RQ2: }
What is the gain in terms of the number of tests reduced?
}

\RQtwo

The effectiveness of our approach is measured by the ratio of number of tests selected over the total number of tests, that is,
\begin{eqnarray*}
Gain = 1 - ( 100 \times |T'| / |T| )
\end{eqnarray*}

\subsection{Results}

\RQone

Table~\ref{tab:missed-mutations} shows the number of mutations generated for each case study in the first column. The second column shows the mutants killed by the whole test-suite. The third column shows how many mutants were missed by the reduced test-suite.

The results confirm that all mutants that were killed by the whole test-suite, were also killed by the selected test-suite. The only exception was \textit{Apache Commons Collections} where 3\% mutants  survived. The detailed analysis showed that PIT added some mutations to the static initializers in the Java code. The fact that the \jvmsniffer only traces method calls, the tests that cause those blocks to be executed were not selected. When adding those tests manually, the number of mutants missed dropped to zero.

\begin{table*}
	\caption{Comparison of mutation coverage.}
	\label{tab:missed-mutations}
	\begin{center}
		\begin{tabular}{llll}
			\hline\noalign{\smallskip}
			\textbf{Case Studies  }             & \multicolumn{1}{l}{\textbf{\# of mutations}} & \multicolumn{1}{l}{\begin{tabular}[c]{@{}l@{}}\textbf{Killed by}\\ \textbf{whole test-suite}\end{tabular}} & \multicolumn{1}{l}{\begin{tabular}[c]{@{}l@{}}\textbf{Missed by}\\ \textbf{selected tests}\end{tabular}} \\
			\noalign{\smallskip}\hline\noalign{\smallskip}
			Apache Commons Collections & 348                                 & 37\%                                                                                         & 3\%                                                                                           \\
			Apache Commons Lang        & 1360                                & 53\%                                                                                         & 0\%                                                                                           \\
			Eclipse Collections        & 29                                  & 53.5\%                                                                                       & 0\%                                                                                           \\
			Netty                      & 1363                                & 33\%                                                                                         & 0\%                                                                                           \\
			RxJava                     & 329                                 & 72\%                                                                                         & 0\%                                                                                           \\
			Seata                      & 192                                 & 17\%                                                                                         & 0\%                                                                                           \\
			OkHttp                     & 543                                 & 55.67\%                                                                                      & 0\%                                                                                           \\
			MyBatis                    & 169                                 & 84\%                                                                                         & 0\%                                                                                           \\
			UML Reverse Mapper         & 27                                  & 55\%                                                                                         & 0\%                                                                                           \\
			Symja Parser               & 182                                 & 32\%                                                                                         & 0\%                                                                                          \\
			\noalign{\smallskip}\hline
		\end{tabular}
	\end{center}
\end{table*}

\RQtwo

Table~\ref{tab:test-minimization-gain} shows the total number of tests in the whole test-suite and the selected test-suite in the first and second columns respectively. The third column shows the percentage reduction in tests. As per the results in Table~\ref{tab:missed-mutations}, running only the selected test-suite is sufficient for the changes introduced due to mutations. There is a significant reduction in the number of tests averaging $\approx$50\%. Eight out of ten case studies had double digit reduction and a few of them is above 80\%.

The only case study that could not achieve any reduction was \textit{Symja Parser}, which is a math utility library. Due to the nature of functions implemented in this library and the fact that we have used math related operators in PIT, almost all methods were mutated. Therefore, all tests in the whole test suite were selected by our approach.

\begin{table*}
	\caption{Tests minimization gain against the whole test-suite.}
	\label{tab:test-minimization-gain}
	\begin{center}
		\begin{tabular}{llll}
			\hline\noalign{\smallskip}
			\multicolumn{1}{c}{\textbf{Case Studies}} & \textbf{\# of tests} & \textbf{\# of selected tests} & \textbf{Gain} \\
			\noalign{\smallskip}\hline\noalign{\smallskip}
			Apache Commons Collections       & 3900        & 744                  & 80.92\%   \\
			Apache Commons Lang              & 3039        & 1814                 & 40.31\%   \\
			Eclipse Collections              & 178         & 61                   & 87.24\%   \\
			Netty                            & 1213        & 956                  & 21.19\%   \\
			RxJava                           & 3571        & 1917                 & 46.32\%   \\
			Seata                            & 245         & 47                   & 80.81\%   \\
			OkHttp                           & 1241        & 1125                 & 9.34\%    \\
			MyBatis                          & 878         & 617                  & 29.73\%   \\
			UML Reverse Mapper               & 67          & 16                   & 76.12\%   \\
			Symja Parser                     & 83          & 83                   & 0\%     \\
			\noalign{\smallskip}\hline
			\textbf{Average}						&      \textbf{14593}  				& 	\textbf{7380}				& \textbf{49.43\%} \\
			\noalign{\smallskip}\hline
		\end{tabular}
	\end{center}
\end{table*}

\subsection{Threats to Validity}

The case studies used in the experiments are real-life projects, but they are not representative of a large integrated system of components. As with any type of evaluation, additional case studies will be required for gaining further confidence in our approach.
Having said that, our objective was to evaluate the approach for test selection based on code changes, for which, these projects provide a sufficient ground. All of these projects are in active development and widely used in the opensource community. We believe that they provide realistic and diverse examples of what our approach needs to be able to handle in practice. 

One threat to validity might come from the mutation strategy used in our experiments. In order to reduce bias, we used the default settings of PIT's operators for all case studies. We repeated each experiment 3 times and found the same result. This is due to the fact that PIT always generates same mutants for an unchanged code. 
Another related threat is the presence of equivalent mutants. PIT tries best to ignore such mutants using a built-in bytecode matching library and removes mutants that are trivially equivalent. Nevertheless, equivalent mutants do not pose a particular problem in our experiments as we compare results on the same set of mutants.

Another threat may come from the implementation of \jvmsniffer that how accurately it computes the mapping between tests and method executions. The tool uses Java Debug Interface API that has been around since Java 1.4 and quite stable~\cite{JDI}. Moreover, we tested \jvmsniffer on all case studies before the experiments and compared the results with the ones in the call graph generated by \textit{Intellij IDEA}~\cite{Krochmalski2014}. 

\subsection{Limitations}

Our implementation of the approach has some limitations listed below:

\subsubsection{Sequential execution:}
The approach is designed to run tests in a distributed environment, and multiple JVMs are supported. However, tests must execute sequentially, otherwise \textit{JVMSniffer} cannot distinguish which method invocation is caused by which test.

\subsubsection{Method level changes:}
Changes to field initializations, static blocks and constants are currently not considered. This
limitation is not fundamental to our approach as most regression test selection techniques~\cite{Law2003}\cite{Orso2003}\cite{Chianti2004}\cite{Orso2004} are applied on the method level, in contrast to coarser-grained approaches~\cite{Ekstazi_issta_2015}\cite{Celik2017}, which are based solely on Java file/class level changes. 

\subsubsection{External resources:}
Changes to configurations, properties files, third-party dependencies and other external factors like database, environment variables, are not considered in this approach. Generally speaking, changing resources between multiple test runs is not recommended from the perspective of reproducibility~\cite{Frattini2016}.

%

\section{Related Works}
\label{sec:related_works}

We divide the related works into two categories. First, we explore research works aiming for regression test selection. Then, we note some production tools that use change analysis for test selection. 


\subsection{Research Works}

Traditional Regression Test Selection techniques compute detailed program changes by traversing control flow graphs of two versions of a program and their associated tests at different code granularities, such as statement block, method, class, or file levels~\cite{Yoo2012}. Then, the tests which overlap the elements of the graph are selected for execution. Such techniques require a lot of processing and incur non-trivial overhead~\cite{zhang2018hybrid}. Therefore, researchers have also proposed techniques that are applied only on the method level granularity with the combination of static and dynamic analysis.

One of the early works in this realm is \textit{Chianti}~\cite{Chianti2004} that relies on the computation of structural differences between the two versions of a Java program. It computes Abstract Syntax Tree (AST) of the source program, followed by identifying changes using predefined rules. It generates Control Flow Graph (CFG) to associate tests with the method changes and determines a subset of tests relevant to the changes. Our approach is similar to Chianti with respect to applying rules to cater the language specific semantics and the concepts of object-oriented programming (e.g., polymorphism). Our approach is different in how the test selection process works. We do not compute AST or CFG to determine the test association, which could be computationally expensive. We obtain this information dynamically from the test execution and create the mapping (or association) between the tests and the methods. This fits well with the CI/CD methodology where the code merges are frequent, and therefore computing such trees or graphs could have a detrimental effect on the build timescales~\cite{zhang2018hybrid}. In our case, there is no overhead of creating this mapping. The only pre-processing required in our approach is the identification of method changes at the bytecode level (via \textit{Soot} optimization framework~\cite{soot}). We argue that this is a significantly less overhead compared to computing a full CFG of the source code as well as the tests. The main reason is that we use the CFG analysis only if a newly added method has the same method signature as an existing method; so the method override/overload rules, i.e., Rules \ref{rule:override}/\ref{rule:overload}), could be applied.


In contrast to static analysis techniques, dynamic techniques have been used for impact analysis to support regression testing~\cite{Law2003}\cite{Orso2003}\cite{Orso2004}. Orso et al.~\cite{Orso2003}\cite{Orso2004} collect execution traces from program instrumentation. The execution traces consists of all methods that were called during program execution. For each code change, they compute an approximate dynamic slice
based on the execution traces that traverse the change. The impact set is the union of the slices computed for each change.
Separately, they select an initial set of tests that traversed at least one change based on coverage information. 
Then, they use the impact set from the execution traces to assess whether, according to the user supplied information (which the authors called `field data'),
the initial set of tests is adequate, i.e., there exists at least one test that traverses the affected method after traversing the change. This step is based on the intuition that the executions that traverse changed parts of the code are more likely to traverse the affected methods. 
The main difference between their work and ours is that their goal is to determine which affected methods are not tested by the given set of test cases. On the contrary, our aim is to find the test cases that are affected by those changes. Secondly, their technique is not safe as they rely on coverage data from static analysis~\cite{Chianti2004}. In contrast, we collect test mapping from execution traces. Thirdly, they rely on program instrumentation, whereas we do not require any changes to the source code for our implementation. 

Lehnert et al.~\cite{Lehnert2012} used UML state machines to identify impacted test cases. In their approach, a test path is a sequence of state machine transitions, which is used to test the class that corresponds to the state machine. For every method changed, the corresponding transition in the state machine which uses this method in its events, guards, or actions is marked as modified. The test sequences that contain this transition is then selected. The underlying hypothesis is that the changes can be identified through design models as the equivalence between the UML models and the implementation~\cite{Heger2014}. Therefore, they introduced specific rules to perform impact analysis on the changed model elements and to build traceability links between model elements and test cases. Cazzola et al.~\cite{cazzola2022bridging} noted that UML behavioral models such as state machines are often incomplete or quickly outdated, which limits the applicability of existing model-based approaches. As an improvement, they proposed UML class diagrams as a more reliable modeling source, and applied fuzzy logic to deal with inconsistencies. They also rely on test cases modeled as a class diagram, the so-called \textit{call usage dependency relationships} model. 
Similarly, Sun et al.~\cite{sun2010change} used \textit{OOCMDG} --- a dependency graph of classes, methods and variables to create impact rules. The rules identify the impact set according to the change types and then computing the union of all change types for each entity. However, their approach does not work for method body changes and statement changes~\cite{Lehnert2012}. In contrast to all these works, we do not rely on any design models or documentation to identify changes. Moreover, our aim is to select test cases that are already implemented as a code rather than represented as abstract models.

Azizi and Do~\cite{Azizi2018} used information retrieval (IR) techniques to select tests during regression testing. 
They identify common keywords in the difference of two program versions and construct IR queries using the keywords.
They also tokenize the source code of the tests to compute their diversity, and build a graph to represent the relationship among tests annotated with the diversity score. Then, they collect tests whose cosine similarity to the queries is higher than a predefined threshold. Finally, they select the most diverse tests among the collected tests. As normally the case with IR techniques, this solution only works if the source code of the program and tests use mostly the same keywords. The ability to select tests is highly dependent on the textual similarities (adjustable by the cosine similarity threshold) and the ratio of the number of tests to the number of queries. 

Symbolic execution techniques have also been used in regression test selection~\cite{Guo2016}\cite{Pasareanu2013}\cite{Yang2014}. Yang et al.~\cite{Yang2014} leveraged symbolic execution with static analysis to identify program statements relevant to changes in the program. They used static analysis to determine the modified program statements; then only the \textit{affected} parts of the program was explored during symbolic execution. Whilst the goal of this technique is to make symbolic execution efficient in program analysis, Guo et al.~\cite{Guo2016} argued that this may leave out some paths if the affected paths are equivalent to some paths in the previous version. They demonstrated their findings on an example of a concurrent program that is executing two threads, where a change in the code of one thread is not identified in the other thread. Hence, the path is not selected in the symbolic exploration. Our approach is not affected by this shortcoming because the test executing the two threads is selected if a change in the code of any of the threads is detected. Nevertheless, the most important drawback of the symbolic execution approach is the algorithmic complexity that can be very expensive~\cite{Yoo2012}.

\subsection{Tools}

Code coverage tools use some degree of impact analysis to generate test reports. For instance, \textit{junit4git}\footnote{https://github.com/rpau/junit4git} implements JUnit 4/5 extensions to generate reports on tests impacted by the code changes. By using the JVM instrumentation mechanism, the tool can deduce which methods are invoked by which tests. This information is then stored as GIT notes, which is used at the next run of tests. The tool can only apply the impact analysis at the unit level. As noted before, TIA at the unit test level is not very beneficial as they are usually faster to run. The real benefit is filtering integration or acceptance tests.

Google’s internal build system \textit{Blaze}\footnote{https://bazel.build/} associates tests with the production code by maintaining a fine-grained directory tree of their source code. Similar technique is applied in \textit{Buck}\footnote{https://buck.build/} for Android apps from Meta and \textit{Pants}\footnote{https://www.pantsbuild.org/} from Twitter. However, all of them support only \textit{monorepos}~\cite{Brousse2019}, where code for many projects are stored in the same repository. This is akin to performing TIA at the unit level since all tests can be traced back to their relevant methods statically.

Microsoft's Visual Studio has incorporated a TIA approach on the integration level for Azure Pipelines~\cite{pratab2017,vstudio_magazine}. It examines the test execution data to identify tests which call the modified methods of a C\# class. Unfortunately, the details of their work are not publicly available. Furthermore, it only works in a single machine topology where tests and system under test must be running on the same server~\cite{vstudio}. Our approach does not have this limitation thanks to \jvmsniffer\ that is deployed on each server running tests.

\section{Summary and Future Works}
\label{sec:conclusion}

This paper presents an approach for Test Impact Analysis on the integration level, where the source code is  distributed in multiple repositories. The approach computes program changes on the method level granularity statically, and selects the impacted tests based on their association with method calls that are collected dynamically. Additionally, we have proposed rules for method changes, additions and removals to cater for Java language semantics. We have used mutation testing to evaluate our approach on a number of opensource Java projects. First, we ran the mutated programs with the whole test suite, then executed the tests selected by our approach to compute the effectiveness. The results achieved $\approx$50\% of test minimization gain with same mutants killed as with the whole test suite.


Our approach for Test Impact Analysis is based on test selection at the method level, but it can be applied on different levels of granularity of program changes, such as statement, class, file or even module level. A simple solution would be to extend the mapping to store additional information besides method calls for each test. However, coarse-grained levels may select tests more than necessary, whereas finer-grained levels may have more processing overhead~\cite{Ekstazi_issta_2015}\cite{zhang2018hybrid}. We shall investigate multiple granular levels in our approach to find the best trade-off between precision and efficiency in the context of CI/CD pipelines. Another interesting area of research is combining test selection with test prioritization to make delivery pipelines more cost-effective. We are looking into techniques to prioritize the selected tests that are more likely to fail based on historical test executions in the pipeline.

\bibliography{tia}%

\end{document}